\documentclass[usedcolumn,useAMS,usenatbib]{mnras}

\usepackage{graphicx,times}
\usepackage{epsfig}
\usepackage{multirow}
\usepackage{booktabs}
\usepackage{mathptmx}
\usepackage{amssymb,amsmath}
\usepackage{xcolor}
\usepackage{rotating}
\usepackage{gensymb}
\usepackage[normalem]{ulem}

\newcommand{\Msun}{{\rm M}_{\odot}}

\newcommand{\hmpc}{h^{-1}{\rm Mpc}}

\newcommand{\mpc}{{\rm Mpc}}
\newcommand{\kpc}{{\rm kpc}}
\newcommand{\kms}{{\rm ~km~s^{-1}}}

\def \sag{\textsc{sag}}

\def \SAG0{\textsc{sag0}}
\def \SAG1S{\textsc{sag1s}}

\def\apjl{ApJ}
\def\apjs{Astrophys.\ J.\ (Supp.) }
\def\aap{A\&A}

\title[SPs in bulges of MW-like galaxies]{Stellar Populations in a semi-analytic model I: bulges of Milky Way-like galaxies}
\author[I. D. Gargiulo et al.]{I. D. Gargiulo$^{1,3}$\thanks{E-mail: gargiulo@fcaglp.unlp.edu.ar}, S. A. Cora$^{1,2,3}$, C. A. Vega-Mart\'inez$^{1}$, O. A. Gonzalez$^{8}$, M. Zoccali$^{9}$ \newauthor  R. Gonz\'alez$^{4,5}$, A. N. Ruiz$^{3,6,7}$ and  N. D. Padilla$^{4,5}$\\
$^{1}$Instituto de Astrof\'isica de La Plata (CCT La Plata, CONICET, UNLP), Paseo del Bosque s/n, B1900FWA, La Plata, Argentina.\\
$^{2}$Facultad de Ciencias Astron\'omicas y Geof\'{\i}sicas, Universidad Nacional de La Plata, Paseo del Bosque s/n, B1900FWA, La Plata, Argentina.\\
$^{3}$Consejo Nacional de Investigaciones Cient\'{\i}ficas y T\'ecnicas, Rivadavia 1917, C1033AAJ Buenos Aires, Argentina.\\
$^{4}$Instituto de Astrof\'{\i}sica, Pontificia Universidad Cat\'olica de Chile, Av. Vicu\~na Mackenna 4860, Santiago, Chile.\\
$^{5}$Centro de Astro-Ingenier\'{\i}a, Pontificia Universidad Cat\'olica de Chile, Av. Vicu\~na Mackenna 4860, Santiago, Chile.\\
$^{6}$Instituto de Astronom\'{\i}a Te\'orica y Experimental (CCT C\'ordoba, CONICET, UNC), Laprida 854, X5000BGR, C\'ordoba, Argentina.\\
$^{7}$Observatorio Astron\'omico de C\'ordoba, Universidad Nacional de C\'ordoba, Laprida 854, X5000BGR, C\'ordoba, Argentina.\\
$^{8}$UK Astronomy Technology Centre, Royal Observatory, Blackford Hill, Edinburgh EH9 3HJ, UK \\
$^{9}$The Milky Way Millennium Nucleus, Av. Vicuña Mackenna 4860, 782-0436 Macul, Santiago, Chile.\\
}

\begin{document}

\date{Accepted 2017 August 21. Received 2017 July 18; in original form 2017 April 4}

\pagerange{\pageref{firstpage}--\pageref{lastpage}} \pubyear{2014}

\maketitle

\label{firstpage}

\begin{abstract}

We study the stellar populations of bulges of Milky Way-like (MW-like)
galaxies with the aim of identifying the physical processes
involved in the formation of the bulge of our Galaxy.
We use the semi-analytic model of galaxy formation and evolution \sag~adapted 
to this aim; this kind of models can trace the properties of galaxies and their
components like stellar discs, bulges and halos, but resolution
limits prevent them
from reaching the scale of stellar populations (SPs).
Properties of 
groups of stars formed during single star formation events are stored and 
tracked in the model and results are compared with
observations of stars in the galactic bulge. MW-like galaxies are selected 
using two different criteria. One of them considers intrinsic photo-metric 
properties and the second is focused on the cosmological context of the local 
group of galaxies (LG). 
We compare our model results with spectroscopic 
and photometric stellar metallicity distributions. We 
find that $87\%$ of stars in bulges of 
MW-type galaxies in our model are accreted and formed in starbursts 
during disc instability events. 
Mergers contribute to $13\%$ of the mass budget of the bulge
and are responsible for the low metallicity tail of the distribution.
Abundance ratios of $\alpha$ elements with respect to iron, [$\alpha$/\text{Fe}], 
are measured in SPs of model galaxies. The patterns found in the model for 
SPs with different origins help to explain the lack of a gradient 
of [$\alpha$/\text{Fe}] ratios in observed stars along the minor axis of the bulge.  


\end{abstract}


\section{Introduction}
\label{sec:intro}

The problem of the formation of bulges remains current.  Today it is 
well known that mergers and interactions of galaxies play an important role in 
bulge formation \citep{KormendyKennicutt2004}. This 
occurs predominantly at high redshifts due 
to the fact that in a $\Lambda$CDM Universe the relative number of mergers 
is higher at earlier times, while at lower redshifts the merger rate 
decreases and the galaxies evolve secularly \citep{Athanassoula2013}.  
There are different processes that 
drive bulge formation via mergers. 
For example, bulges form by collisionless accretion 
of stellar systems into the bulge \citep{Aguerri2001} or violent 
relaxation of a fraction of the stellar and gaseous components of a disc 
triggered by the accretion of a 
satellite galaxy \citep{Hopkins2009, Zavala2012}. 
On the other hand, bulges can also be 
the consequence of secular evolution. 
During this slow evolution, 
galactic discs with high surface densities can become unstable 
against small perturbations \citep{Efstathiou1982, MoMaoWhite1998} or 
even mergers \citep{Padilla2014}.
These instabilities often produce a bar which drives the 
secular evolution transferring stars and gas of the disc to the central regions, 
where stars are formed in starbursts \citep{Kormendy2013}.
 
Other bulge formation mechanisms are still considered in the literature, like 
the gravitational collapse in-situ of a gas cloud \citep{Pipino2004} 
or the 
formation of giant clumps originated from gas in discs of high 
redshift galaxies that reach the galactic center due to dynamical 
friction and form stars in bursts \citep{Dekel2009, Ceverino2010, Perez2013}. 
In this complex context, the case of the Milky Way (MW) bulge is our best 
chance to shed some light to the problem of bulge formation.

The chemical and structural properties of 
stellar populations (SPs)
hold the memory 
of their formation and evolution.
A common observational approach to gain insight into the formation processes 
that a composite SP has faced is to study the chemical 
signatures imprinted in their stars. This method is commonly known as 
'{\it stellar archaeology}'.  
Important efforts have been carried out 
 to observe the bulge of the MW in the near-infrared, like the Vista Variables 
in the V\'{\i}a L\'actea survey \citep[VVV,][]{Minniti2010}, where the 
extinction effects are much lower than in optical wavelengths.
Now we can observe stars individually and have access to all the information 
that they keep in their properties. In the last years, as a result of the
large amount of measurements of the metallicity and kinematics of bulge 
stars, much discussion has been given in relation with the origin of the
bulge of our Galaxy.    
Although a large fraction of the light in infrared is contributed by red
giant stars, it is possible to observe stars in the turn-off of 
color-magnitude diagrams by means of a technique that employs gravitational 
microlensing.  Using 58 microlensed dwarf and subgiant stars, 
\citet{Bensby2013} conclude that there might exist a combination of 
multiple SPs by constructing the metallicity distribution 
function (MDF) of the stars. \citet{Ness2013} also point out that the MDF of 
stars in the bulge can be described by at least 3 different subpopulations. 
More recently, \citet{Zoccali2016} construct iron distribution functions (IDF)
for a set of fields in the GIRAFFE Inner Bulge Survey(GIBS) and conclude that 
the bulge has two different metallicity components.

The chemical abundances of SPs are a fundamental 
tool to validate and contrast the different bulge formation hypothesis.  
Semi-analytic models of galaxy formation can provide large numbers 
of galaxies similar to the Milky Way and predict the chemical enrichment of
SPs.
One of the first semi-analytic models of galaxy formation 
that implemented a detailed chemical enrichment model was presented 
by \citet{Cora2006}. This model considers the contribution of different
chemical elements generated by core-collapse supernovae (SNe CC)
and supernovae type Ia (SNe Ia), as well as by low- and intermediate-mass 
stars, relaxing the instantaneous recycling approximation hypothesis.
A similar chemical implementation was introduced by \citet{DeLucia2014} in their
semi-analytic model of galaxy formation to study the abundances of single 
SPs formed in different components of galaxies in halos of the 
Aquarius simulation \citep{Springel2008}. 
They find a good
agreement between the iron content of disc populations in the model with 
observed abundances of stars in the Galactic disc. However, the abundances 
of SPs in the modeled bulge component is shifted low with respect to 
observations, likely due to their decision of switching off the disc 
instabilities effects in their 
model that would contribute to the bulge formation.  
Our goal is to get insight into the formation mechanisms of bulges 
in galaxies like the Milky Way, that is, to understand the origin of the 
stars that form the bulge, by making use of a semi-analytic model
of galaxy formation based on the one presented by \citet{Cora2006}
with the implementation of a novel approach that allows to reach 
the scale of SPs in simulated galaxies.
This paper is organized as follows. Section \ref{sec:model} describes the 
central aspects of the simulation and semi-analytic model used throughout 
this work and presents the details of the calibration of the model,
i.e., the tuning of its free parameters.
Section \ref{sec:MW} describes 
two different criteria applied to select MW-type galaxies from our
simulated galaxy catalog;
we analyze the sample that emerges from their combination.
In Section \ref{sec:results}, we present the results 
obtained from our investigation. Finally, in section \ref{sec:discandconc} we 
discuss the results and present our conclusions.  

\section[]{Model of galaxy formation}
\label{sec:model}

We use the semi-analytic model of galaxy formation and evolution
\sag~(acronym for {\it Semi-Analytic Galaxies})
 presented by \citet{Cora2006}, and further improved in the 
last years through the implementation of feedback from Active 
Galactic Nuclei (AGN) \citep{Lagos2008},
the refinement of the characterization of galactic 
discs \citep{Tecce2010} and more recently, 
the inclusion of improved treatment of bulge 
sizes \citet{munnozarancibia2015}, and of extended
bursts of star formation along with new stellar yields and different 
initial mass function (IMF) 
configurations \citep[][ hereafter G15]{Gargiulo2015}. 
Cora et al. (2017, in prep.) 
present the most updated version of \sag~that considers environmental effects as a result
of ram pressure and tidal stripping, and   
a new model for supernovae feedback. 
For a more complete description of the model, we refer the reader to 
the latter work. 
In Section~\ref{sec:BulgeFormation}, we highlight the aspects of \sag~that are
more relevant to the current investigation,
indicating the changes introduced for this particular study
and
presenting the method employed to tune the free parameters of the model
in Section~\ref{sec:Calibration}.
In Section~\ref{sec:SPs}, we describe the way in which SPs
and their chemical properties are tracked.

We run the semi-analytic model on top of the dissipationless DM only 
{\em N}-body simulation {\rm MDPL2}, part of the {\rm MULTIDARK} 
project \citep{Prada2012}. 
MDPL2 simulation follows the evolution of $3840^3$ particles 
in a cubic box of comoving sidelength
$L=1\,h^{-1}{\rm Gpc}$ \citep{klypin16}
using Planck cosmological parameters \citep{PlanckColl16}.
The simulation was evolved starting from $z_{\rm ini}=120$
to the present epoch, storing 126 outputs 
equally spaced in $\log_{10}(a)$ 
between $z=17$ and $z=0$.
DM haloes were identified using {\rm ROCKSTAR}~algorithm
\citep{Behroozi2013}, which allows to select self-bound substructures
(subhaloes) having at least 20 particles, with a mass of
$3\times 10^{10}\,{h^{-1}}\,\Msun$ each.
The merger trees of the halos were generated using 
the {\sc ConsistentTrees} code, which  
establishes the evolution of the halos by explicitly ensuring the 
consistency of halo properties through the snapshots of the simulations
\citep{Behroozi2013}.

The complete volume of the MDPL2 simulation is composed by a large amount 
of data. 
For this reason, in order to ease the analysis, in this work we use only 
several samples of haloes extracted from the full simulation.
 The selection and extraction of this samples has been achieved by using
the {\it forests} of the simulation. In this context, a forest is defined 
as a group
of halos related either by spacial or historical links, so that a forest can be
composed by one or more merger trees. According to this, each forest 
corresponds to the minimum structure which can be analysand in isolation.

\begin{table}
\small

\caption{Best-fitting values of the free parameters of the model.}
\centering
\begin{center}
\begin{tabular}{l c c l}
\hline
 parameter & value \\ 
\cline{1-2}
$\bf{\alpha}$ & 0.017 \\ 
$\bf{\epsilon}$ & 0.33 \\
$\bf{\epsilon_\text{ejec}}$ & 0.02 \\
$\bf{\gamma}$ & 0.16 \\
$\bf{f_\text{BH}}$ & 0.03 \\
$\bf{\kappa_\text{AGN}}$ & $5.76\times10^{-5}$ \\
$\bf{f_\text{hot,sat}}$ & 0.3 \\
$\bf{f_\text{pert}}$ & 50.0 \\
$\text{f}_\text{cold,DI}$ & 0.75 \\
\hline

\end{tabular}
\\

{\bf Note.} All parameters were used for calibration of the model and are described 
in section Sec.~\ref{sec:Calibration}. 

\end{center}

\label{table1}
\end{table}

\subsection{Bulge formation mechanisms}
\label{sec:BulgeFormation}

In the \sag~model, bulges
are formed via mergers and global disc instabilities (DI). 
In the following we describe the conditions involved in these
two processes and the way in which the baryonic components
of galaxies are affected by them. While the treatment of mergers
is the same as the one implemented in previous versions of \sag,
many aspects of DI have been modified for the purposes of the present study.
Both mechanisms trigger starbursts that 
involve a timescale that regulates the consumption of the cold gas involved
in the process as implemented by \citet{Gargiulo2015}, 
instead of being instantaneous as in former versions of \sag. 
We refer the reader to their work for more details 
about the treatment of extended bursts.

\subsubsection{Galaxy mergers}
\label{sec:mergers}

Galaxy mergers can be major or minor. When a merger occurs, the stellar
mass ratio between the satellite galaxy and the central galaxy, $M_\text{sat}/M_\text{cen}$,
is evaluated. If $M_\text{sat}/M_\text{cen} > 0.3$ the merger is considered a major one. 
In this case, all the gas in the remnant galaxy is consumed in a starburst
and contributes to the bulge formation. The whole stellar disc is relaxed and
transferred to the bulge. If the merger is minor ($M_\text{sat}/M_\text{cen} \le 0.3$)
a starburst is triggered when the fraction of cold gas 
 with respect to the total mass  present in the 
disc of the central galaxy, $M_\text{cold,cen}/M_\text{disc,cen}$, 
is larger than a fixed
parameter $f_\text{burst} = 0.6$ \citep{Lagos2008}. In this case, 
the perturbation
introduced by the merging satellite drives all the cold gas from both
galaxies into the bulge component, where the stars form in a starburst. If 
this condition is not fulfilled, or the ratio $M_\text{sat}/M_\text{cen} \le 0.05$, 
the minor merger is dry, and only the stars of the merging 
satellite galaxy
are transferred to the bulge component of the central galaxy.

\subsubsection{Disc instabilities}
\label{sec:DI}

Usually, semi-analytic models adopt
the  Efstathiou Lake \& Negroponte \citep[ELN,][]{Efstathiou1982} criterion
to model the disc instabilities in galaxies\citep{Benson2010, Cattaneo2017,
Gonzalez-Perez2014}.
 This criterion establishes that a galactic disc becomes bar unstable 
when a galactic disc is sufficiently massive that its self-gravity 
is dominant.
This happens 
when
\begin{equation}
\epsilon_{\rm d} \equiv \frac{V_{\rm disc} }{ (G M_{\rm disc} 
/ R_{\rm disc})^{1/2}} \le \epsilon_{\rm thresh},
\label{eq:diskinstab}
\end{equation}

\noindent where $M_{\rm disc}$ is the mass of the disc,  
$R_{\rm disc}$ is the disc scale radius, 
and $V_{\rm disc}$ is the circular velocity of the disc. 
In our model, $V_{\rm disc}$ is given by the velocity where the rotation curve flattens, 
which approximate by the velocity calculated at $\sim 3\,R_{\rm disc}$ 
\citep{Tecce2010}. 
Both $R_{\rm disc}$ and $V_{\rm disc}$ are estimated by
using the fitting formulae provided by \citet{MoMaoWhite1998}.
Gas and stars are accounted in
the disc mass.  
We adopt $\epsilon_{\rm thresh}=1$.

The inclusion of the disc instability
process to model the galactic bulges in a semi-analytic model of galaxy 
formation is not undisputed. 
\citet{Athanassoula2008} presented results of {\it N}-body simulations 
of discs embedded in dark matter haloes with different values for the 
ELN threshold. 
They show that several discs with $\epsilon_{\rm thresh} < 1.1$ 
remained stable against bar formation.  
\citet{DeLucia2014} also argued against the inclusion of this process when 
studying the metallicity distributions of stars from the different components 
of high resolution MW sized galaxies. However, this model, although 
not completely accurate, proves to be 
essential in order to obtain suitable morphologies in galaxies in  
semi-analytic models \citep{Lagos2008, Parry2009, Guo2011}.
The main criticism to be made to the ELN criterion is that the original 
simulations from where it was derived were two-dimensional and, hence, 
considered a rigid halo hosting the disc. This does not allow the transfer 
of angular momentum between the stars and the particles of the DM Halo. 
The angular momentum exchange between the DM halo and the disc seems
to be an important effect in order to maintain the stability for 
discs \citep{Athanassoula2002}, among others, like the degree of random 
motions in the disc \citep{AthanassoulaSellwood1986} and interaction with 
other galaxies in a cosmological context \citep{Scannapieco2012}.

To overcome these problems, we propose here a new combination of criteria 
for galaxies to become unstable in our semi-analytic model. 
We follow the results of \citet{Algorry2017}, who studied barred galaxies 
in the EAGLE simulations \citep{Schaye2015, Crain2015}. The EAGLE simulations 
are a suite of state of the art hydro-dynamical cosmological simulations with 
an unprecedented combination of resolution and size that allowed the authors
to derive properties of barred galaxies as a population in a cosmological 
context. They conclude that, in order to develop strong bars, discs
must be locally and globally dominant. Therefore, they propose a new criterion 
that takes into account the 
global gravitational importance of the whole system, 
in addition to the local 
gravitation of the disc, to explain the proportions of barred galaxies in 
the simulations. They estimate such global gravitational importance with: 
\begin{equation}
f_{\rm dec} = \frac{V_{\rm 50} }{ V_{\rm max,halo} },
\label{eq:fdec}
\end{equation}
\noindent where $V_{\rm 50}$ is the circular velocity at half mass radius and 
$V_{\rm max,halo}$ is the maximum circular velocity of the surrounding halo.
If $f_{\rm dec} > 1$, the disc velocity curve declines and the disc is 
gravitationally dominant in the system. 
We compute this quantity in our model replacing $V_{\rm 50}$ and
$V_{\rm max,halo}$
by the circular velocity of the disc, $V_{\rm disc}$,
and the virial velocity of the 
substructure that hosts the galaxy, $V_{\rm vir}$, respectively. 
Thus, in the current version of \sag,
a galaxy is considered able of unstabilizing if it satisfies 
simultaneously
both the ELN criterion 
and the \citet{Algorry2017} criterion, that is, 
if $\epsilon_{\rm thresh} < 1$ and
$f_{dec} > 1$. 

%
%
As in previous versions of \sag, we keep the requirement
of the presence of a neighbouring galaxy to perturb the disc 
and trigger 
DIs. We consider that this occurs when
the mean separation between galaxies in a
main host halo
is smaller than $f_\text{pert}\,R_{\rm disc}$,   
where $f_\text{pert}$ is a 
free parameter of the model.

Stars and gas are transferred from the disc to the bulge when
a DI is triggered. The way in which this mass transfer takes place
is modified in the present work. Instead of allowing 
all the stars and cold gas in the disc to contribute to the formation of the 
bulge component, where the available gas is consumed in a starburst,
bulge formation proceeds in a more gradual way now.
If the disc becomes unstable 
and a perturbing galaxy triggers a DI, 
the necessary amount of mass in stars is transferred to the bulge
in order to achieve stability. 
If this mass transfer is not enough to reach stability, 
a fraction $f_{\rm cold,DI}$ of the cold gas
is transferred and it is consumed in a starburst.
Only during the first instability event suffered by a galaxy, all the
cold gas in the disc is transferred to the bulge component 
($f_{\rm cold,DI}=1$).
This is motivated by simulations
of \citet{Fanali2015}, who found that at the onset of a bar, 
a major episode of
gas infall ocurrs.
The fraction of gas transferred in successive DI events that takes place
before the galaxy reaches stability again is less than unit
and it is considered as a free parameter of the model. 
This gradual transfer of mass allows bulges 
to be formed
and limit the formation of elliptical galaxies by the instability mechanism,
because rarely the whole mass of the disc is needed to be transferred to the 
bulge to stabilize the galaxy.

\subsection{Mass recycling scheme}

Stellar evolution processes give place to stellar winds and supernova
explosions that contribute with newly synthesized material to the surrounding
medium. The fate of this recycled mass that will further contribute
to the formation of a new generation of stars
is highly uncertain. 
Semi-analytic models
of galaxy formation implement this physical process in many different ways.
In previous versions of \sag, the fate of the recycled mass
depends also on the amount of mass reheated 
by SN feedback, so that if the former is smaller than the latter, 
all the recycled mass is deposited in the hot phase, but
a fraction of it is kept in the cold gas otherwise.
We simplify this implementation in the current work considering 
that the amount of recycled mass that will be injected into the
cold gas phase is proportional to the cold gas fraction, that is,
\begin{equation}
{\rm M}_{\rm rec} = \frac{M_{\rm coldgas}}{(M_{\rm coldgas}+M_{\rm disc}+M_{\rm bulge}) } ,
\label{eq:diskinstab}
\end{equation}

where $M_{\rm coldgas}$ is the cold gas mass, 
and $M_{\rm disc}$  and $M_{\rm bulge}$ are the mass of stars
in the disc and bulge, respectively.
The rest of the recycled mass is ejected to the hot phase.
This new prescription allows to recover a better behaviour of
the mass-metallicity relation, being less steep at low stellar
masses than the one obtained 
in former versions of \sag~(Collacchioni et al., in prep.).

\subsection{Chemical enrichment model}
\label{sec:ChemicalModel}

The chemical model implemented in the current version of \sag~is
detailed in \citet{Cora2006}. 
The model tracks the circulation of metals between the different
baryonic components. It
follows the production of eleven chemical elements
(H, $^4$He, $^{12}$C, $^{14}$N, $^{16}$O, $^{20}$Ne, $^{24}$Mg, $^{28}$Si,
$^{32}$S, $^{40}$Ca, $^{56}$Fe)
generated by stars
in different
mass ranges.
As described in \citet{Gargiulo2015}, we adopt a new set of stellar yields
given by  \citet{Karakas2010}, \citep{Hirschi2005} and \citet{Kobayashi2006}
for low and intermediate-mass stars (mass interval $1-8 \Msun$), 
for the mass loss of pre-supernova
stars (He and CNO elements), and for the explosive nucleosynthesis (SNe CC), 
respectively, all of them corresponding to solar metallicities.
These yields are selected to be in
accordance with the large number of constraints for the Milky Way
\citep{Romano2010}.

The model also includes ejecta from SNe Ia.
We assume the 
single degenerate model \citep{Greggio83, Lia2002}
to estimate the SNe Ia rates.
This model assumes that
SNe Ia occurs by carbon deflagration in C-O
white dwarfs in binary systems composed by low and intermediate-mass stars.
Thus, SNe Ia rates depend on the fraction of these binary systems, 
$A_{\rm bin}$. This parameter has impact
on the IDF predicted by the \sag~model (see Sections~\ref{sec:Calibration} 
and~\ref{sec:IDF}), since 
SNe Ia 
contributes with a high amount of iron  
($\sim 0.6\, {\rm M}_{\odot}$).
We adopt the nucleosynthesis
prescriptions from the updated model W7 by \citet{Iwamoto99}.

As we mentioned in the Introduction, \sag~model does not apply the instantaneous
recycling approximation. 
The return time-scale of mass losses and ejecta from all sources considered
are taken into account by our model,
using the stellar lifetime given by \citet{Padovani93}.
This is an essential ingredient for the current study that 
focuses in both the iron and $\alpha$-elements abundances of
stars in the bulge; these products are generated by two distinct
types of SNe whose progenitors are characterized by quite different lifetimes.

\subsection{Calibration of \sag}
\label{sec:Calibration}

The efficiency of the physical processes involved in the \sag~model
are regulated by free parameters.  To calibrate the values of these free parameters, 
we select a set of observational constraints that we expect the model 
to reproduce and apply
the ``Particle Swarm Optimization'' (PSO) technique \citep{Ruiz2015}.
The observational constraints used
to calibrate the current version of 
\sag~are 
\textcolor{blue}{taken from}
a set of constraints defined during the Cosmic CARNage 
workshop\footnote{http://users.obs.carnegiescience.edu/abenson/
CARnage.html} aimed to compare different galaxy formation models
based on the same cosmological DM-only simulation,
as an extension of the comparison project described in \citet{knebe15}. 
These are the stellar mass function at 
$z=0$, for which we use the compilation from \citet{henriques_mcmc_2015}, the star formation rate distribution function 
\citep{gruppioni15}, the fraction of mass in cold gas as a function of 
stellar mass \citep{boselli14}
and the relation between bulge mass and the mass of the
central super massive~BH 
\citep{mcconnell_bhb_2013, kormendy_bhb_2013}.     
The parameters choose to vary in the calibration process of the model are nine,
namely
the star formation efficiency 
($\alpha$), the efficiency of SN feedback from stars 
formed in both the disc and the bulge
($\epsilon$), the efficiency of ejection of gas from the 
hot phase ($\epsilon_\text{ejec}$)
and of its reincorporation ($\gamma$), 
the growth of super massive BHs and efficiency of AGN feedback 
($f_\text{BH}$ and $\kappa_\text{AGN}$, respectively),
the factor involved in the distance scale of
perturbation to trigger disc instability events
($f_{\rm pert}$), 
the fraction that determine the destination of 
the reheated cold gas, $f_\text{hot,sat}$,
and the fraction of gas transferred to the bulge
during DIs, $f_{\rm cold,DI}$.
Most of
the equations that involve these parameters can be found in \citet{Ruiz2015}
and Cora et al. (2017, in prep.).

Once the observational constraints and parameters to vary are chosen,
the best-fitting values
of the free parameters of the model are found by means of the PSO technique,
which is applied to a subsample of 
forests extracted from a smaller box of the
whole MDPL2 simulation that constitute a representative sample
of all the forests contained in the $1\,h^{-1}{\rm Gpc}$ side-length box.
Table \ref{table1} shows the values
of the final set of parameters.
Other parameters were kept fixed during the calibration process, like
the fraction of binary stars involved in the estimation
of SNe Ia rate ($f_\text{bin}$) and 
the exponent that regulates
the redshift evolution of the mass loading factor of the reheated and 
ejected mass, which is fixed in $1.3$ as suggested by \citet{Muratov2015}
(see Cora et al., in prep.).

Although originally we chose $f_\text{bin} = 0.05$, we
allow that parameter to take a larger value
in order to obtain SPs with higher level of iron abundances
that approaches the observed IDF.
Therefore,
the results presented here are obtained with $f_\text{bin} = 0.2$.
This value does not foster a perfect match,
as discussed in Section~\ref{sec:IDF}, but it is the highest
one we can take without affecting significantly other galaxy properties.
In the current version of \sag~we 
adopt a Chabrier IMF. 
 
Details of the calibration  
procedure
and results of the general properties
of the galaxy population generated with the calibrated model
are presented in Cora et al. (2017, in prep).
The changes introduced here regarding the treatment
of the DIs and the fate of the recycled fraction do not modify significantly
the behaviour of those properties, but allows to 
obtain SPs with higher level of iron abundances
that approaches the observed iron distribution functions, as shown in 
Section~\ref{sec:IDF}.
\\

\subsection{Tracking of stellar populations}
\label{sec:SPs}

We include the follow up of SPs in \sag.
In each star formation event, 
we store properties like mass of stars 
formed and masses in individual chemical elements
of the newly formed stars.
We discriminate the information corresponding to the
galaxy stellar components: disc, bulge and halo. 
Because of memory handling issues, we cannot store the totality of the 
formed SPs, which corresponds to all the star formation
events in the evolution of all galaxies. Therefore,
for each MW-like galaxy
we compute the average 
of the stellar mass formed 
and of the corresponding mass of metals locked in the bulge stars,
in every snapshot of the
underlying DM-only simulation. 
Hence, we run \sag~on a selected set of forests contained in the
 $1\,h^{-1}\,{\rm Gpc}$ side-length box of the MDPL2 simulation 
that lead to DM-halos at $z=0$ that host
a MW-like galaxy according to the criteria described in Section~\ref{sec:MW}.
Then, we stack
all the SPs properties in a single compilation of SPs, 
representative of the MW-like galaxies in the simulation. This approach
allows us to handle a higher number of SPs and reduce 
the statistical noise. The inclusion of the tracking of
properties of SPs in a semi-analytic
model of galaxy formation is 
a novel tool, very useful to study the process of formation of 
different galaxy components, in particular the one in which we are interested now, i.e., the 
bulge. This new approach allows, for example, to study how
SPs of different origins are distributed in the chemical abundance space
from a theoretical point of view, helping to understand current 
observations.   
 
\section{MW-like galaxies}
\label{sec:MW}

MW-like galaxies analyzed in this work are those
that fulfill simultaneously  
two independent criteria. On one hand they have to be hosted
by halos that are part of a pair of halos immersed in a group with 
characteristics similar to our Local Group (LG). We refer to them as 
LG analogs. On the other hand, galaxies within these halos must satisfy 
certain morphology and luminosity properties. In this section, we describe 
these two criteria and present the resulting set of MW-like galaxies.

\subsection{LG analogs}
\label{sec:LG}

Galaxies in different environments have dissimilar 
properties. In particular, the mass growth history of galaxies depend on 
the environment in which the galaxy resides, and then the properties  
of their discs and bulges can be affected \citep{LacknerGunn2013}. 

LG analogues are identified by several constraints. First, we search for
host halo pairs in which both members have masses in a wide range of
masses from  $M_{200c}=1 \times 10^{11} \Msun $ to $ 1 \times 10^{13}
\Msun$ $\,$ and are separated by  $0.5-1.3 \mpc$.

Second, to select pairs in relative isolation and to avoid pairs in
triplets or larger groups, we define a quantitative isolation criterion
using the force constraint $F_{i.\rm com}<\kappa F_{12}$, where $F_{i,\rm com}$ is the gravitational force between the pair and any neighbour
 halo $i$ within a $5 \, \hmpc$ radius of the pair center-of-mass,
 $F_{12}$ is the force between the pair, and $\kappa$ is a constant
parameter. The isolation criterion becomes increasingly strict for
decreasing values of $\kappa$. The Milky Way and M31 do not have massive 
neighbours 
within $5\ \mpc$, and  should thus have $\kappa<0.1$. The
actual value of $\kappa$ is, however, uncertain, and  we use a more
relaxed value of $\kappa=0.25$ based on previous tests reported in
\citet{Gonzalez2013, Gonzalez2014}.

The third selection criterion is intended to mimic the absence of massive
clusters in the immediate vicinity of the Local Group. We require that
halos in the LG sample have no neighbour 
halo with mass  $M_{200c}>1.5
\times 10^{14}\ \rm M_{\odot}$ within $12$ Mpc. The mass and distance
limits are somewhat lower than the actual values for the Virgo Cluster.

Fourth, we impose galactocentric radial velocity, tangential velocity, and
pair separation constraint based on literature, but with errors amplified
to $V_{\rm RAD}=-109.3\pm80 \kms$, $V_{\rm TAN}<65 \kms$ and $\Delta
r=770\pm100 \kpc$, respectively, to get 
a sample with more relaxed conditions. 
More details about the LG identification, and reference values can be
found in \citet{Gonzalez2014}.

We apply the criteria with different levels of precision. Finally,
we selected galaxies that lie within $3\sigma$ of the constraints.   
Only $3159$ pairs in the {\rm MDPL2} simulation survive the four criteria.

\subsection{Luminosity and morphology criterion}
\label{sec:M11}

The selection of LG analogs presented in section~\ref{sec:LG} is aimed
to detect galaxies that share a similar cosmological history than the 
MW and M31. Many physical processes experienced by galaxies during their
evolution are affected by their history of mergers and the density of the surrounding
Universe. However, the 
final destiny of the galaxy is not fully determined by these events. 
A number of galaxies of the selected
LG analogs in our simulation do not end with the expected morphology 
or luminosity of the MW. Therefore, we add another constraint to
the selected LG analogs in order to obtain 
the final set of MW-like galaxies in our model.     

We follow the criterion defined by \citet[][M11 hereafter]{Mutch2011}.
They define a galaxy as a MW when three conditions are fulfilled. 
First, they select the most massive galaxy in the halo that contains it. 
In the simulation, these are the central galaxies of a 
main host halo, named as {\em type 0} 
galaxies. Second, the galaxies must have masses in the 
range $10.66 < log_{10}(M_\text{star}[\Msun]) < 11.12$. Third, the galaxies must
show an approximate Sb/c morphology. This is translated to the simulation
by selecting galaxies with a luminosity ratio 
such that the difference between the magnitude of the bulge and of the total
stellar mass in the {\em B}-band is comprised within the range
$1.5 < M_B^\text{bulge} - M_B^\text{total} < 2.6$.
  
\begin{figure}
   \includegraphics[angle=270, scale=0.37]{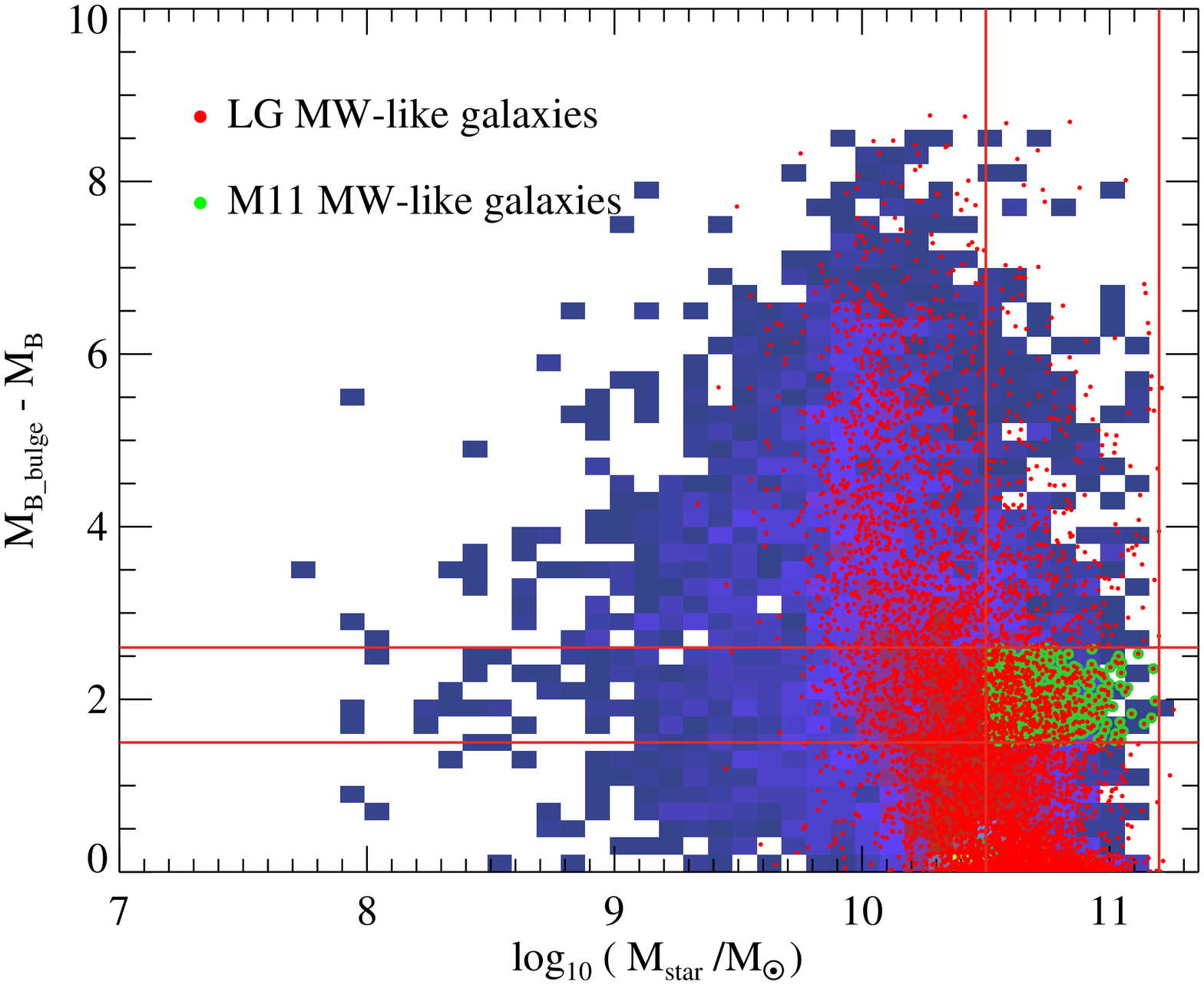}
   \includegraphics[angle=270, scale=0.37]{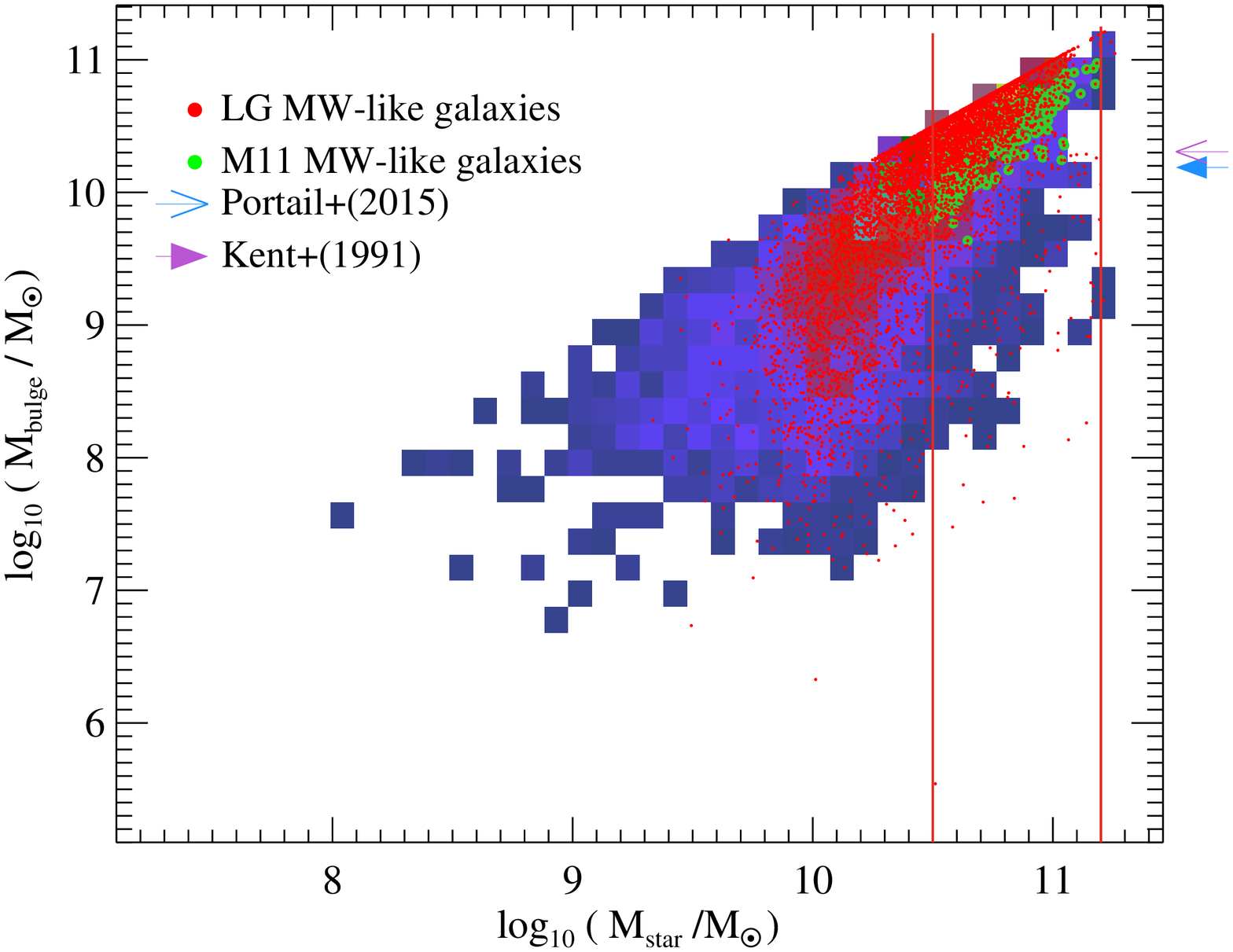}
\caption{{\em Upper panel:} Magnitude difference (ratio of luminosities) in 
{\em B}-filter as a function of stellar mass of galaxies in \sag. Delimited by 
red lines, we show the region in which MW-like galaxies should lie according
to the criterion defined by \citep{Mutch2011} (M11 criterion).
Red points indicate the MW-like galaxies in LG analogs. Green 
points represent the galaxies that fulfill the LG analog and the M11 
criteria simultaneously.
{\em Lower panel:} Mass of the bulge as a function of the stellar
mass of MW-like galaxies. Galaxies selected with the LG criterion are indicated
with red points, and galaxies that also satisfy the M11 
criterion are indicated with green points. Red lines indicate the limit in 
stellar mass adopted in the M11 criterion. The location in both 
$M_B^\text{bulge} - M_B^\text{total}$ vs. $M_\text{star}$  and $M_\text{bulge}$ vs. $M_\text{star}$ planes of all galaxies
that belong to the same forests of the MW-like galaxies 
are represented by a density map in blue. 
The arrows point to the values of bulge mass obtained by different 
authors as indicated in the key (see text).}

\label{fig:SPLGM11}
\end{figure}

In the upper panel of Fig.\ref{fig:SPLGM11}, we show the regions occupied by galaxies 
selected by the two criteria in a magnitude difference 
(or ratio of luminosities) versus stellar mass plot. Red points are galaxies
selected by the LG criterion, while green points are galaxies that fulfill
simultaneously the LG and M11 criteria.
The red
lines show the restricted area imposed by the baryonic 
criterion (M11 criterion). The mass of 
the majority of MW galaxies selected by the LG criterion is below
the lower limit of the M11 criterion, $log_{10}(M_\text{star}[\Msun]) = 10.66$, or have a larger bulge ($M_B^{bulge} - M_B^{total} < 1.5$). Only 612 of 
the selected LG galaxies end with the desired luminosity and morphology
of a MW-like galaxy.
In the lower panel of Fig.\ref{fig:SPLGM11}, we show the mass of bulges of 
MW-like galaxies as a function of stellar mass. The color code is the same
as in the upper panel. The red lines delimit the
range in stellar mass demanded by the M11 criterion. We can see that
the masses of bulges of galaxies that fulfill both the LG and M11 criteria 
are between $9\times10^{9} \Msun$ and $7\times10^{10} \Msun$. 
The arrows represent different estimations of the dynamical mass of the 
galactic bulge. In the background, we include
the density of galaxies that 
belong to the same forests of the MW galaxies in a blue 
palette. 
 
We assume that the stellar mass of the bulge in our model is a good estimation
of the dynamical mass, given that the stellar component of the bulge dominates
the mass budget in the central region of the MW.  We find that the 
stellar bulges of the MW-like galaxies generated by the model
have stellar masses that are consistent to the dynamical
masses estimated in the literature. \citet{Kent1992} modeled for the first 
time the galactic bulge and, 
for the parameters obtained for his model, one can infer that 
the bulge of the MW has a dynamical mass of $1.8\times 10^{10} \Msun$. 
\citet{Sofue2009} also report a dynamical mass of $M_{\rm bulge}=1.8\times10^{10}$.
\citet{Portail2015} found a total mass of $M_{\rm bulge}=1.84\pm0.07\times10^{10}$ by combining dynamical
models and 3D density of Red Clump Giants. They also
estimate a stellar mass of $M_{\rm b, stellar}=1.25-1.6\times10^{10}$.

\subsection{Star formation histories of MW-like galaxies}

\begin{figure}
   \includegraphics[scale=0.46]{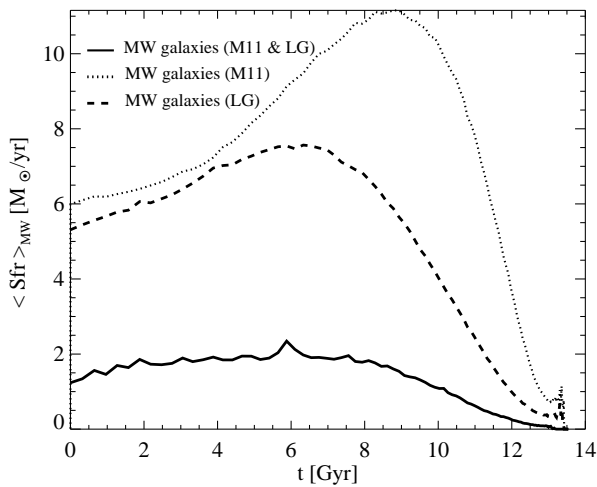}
\caption{Star formation histories of  
MW-like galaxies selected by different criteria as indicated in the keys.
}
\label{fig:SfrMW}
\end{figure}

The total SFR of our galaxy at present time has proven
 difficult to determine accurately due to uncertainties in 
the distance measurements, the IMF assumptions and different 
stellar population synthesis model
(SPS) used. \citet{ChomiukPovich2011}
compiled estimates of the MW SFR in the literature up to 2011
and re-normalized them to a single Kroupa IMF \citep{Kroupa2003}      
and Starburst99 SPS model \citep{VazquezLeitherer2005}. They found
that most of the measurements are consistent, after renormalization, with
a star formation of $\dot{M}=1.9 \pm 0.4 \Msun {\rm yr}^{-1}$.
\citet{Licquia2015} used a bayesian based method to estimate the SFR of the
whole galaxy and found a value of 
$\dot{M}=1.65 \pm 0.19 \Msun {\rm yr}^{-1}$.
In Fig.~\ref{fig:SfrMW} we show the star formation histories of 
MW-type
galaxies selected by the two selection criteria  
described above,
applied both separately and simultaneously. 
The dotted line correspond to galaxies selected with the M11 criterion 
. Galaxies with morphologies resembling the MW show a star formation rate that
peaks at a lookbacktime of 9 {\rm Gyrs} and reach a SFR of 
$\dot{M}=6 \Msun {\rm yr}^{-1}$ 
at z=0. Galaxies in local groups analogs  
are represented by a dashed line and follow a weaker star formation history 
with a peak at 6 {\rm Gyrs} of lookback-time and a SFR at z=0 
of $\dot{M}=5.4 \Msun {\rm yr}^{-1}$. 
The solid line represents the star formation history of galaxies that fulfill
both criteria. We can see that galaxies that lie in the local group analogs and
present a MW like morphology and luminosity are a particular selection of both 
subgroups of 
MW-like galaxies, that prefer a quiet star formation activity 
during their evolution. The average star formation for these galaxies 
at $z=0$ is $\dot{M}=1.26 \Msun {\rm yr}^{-1}$, 
a value that lie inside the 2-sigma 
uncertainties of the observed 
values 
mentioned
previously.

\subsection{Instability of MW-like galaxies}

\begin{figure}
   \includegraphics[scale=0.46]{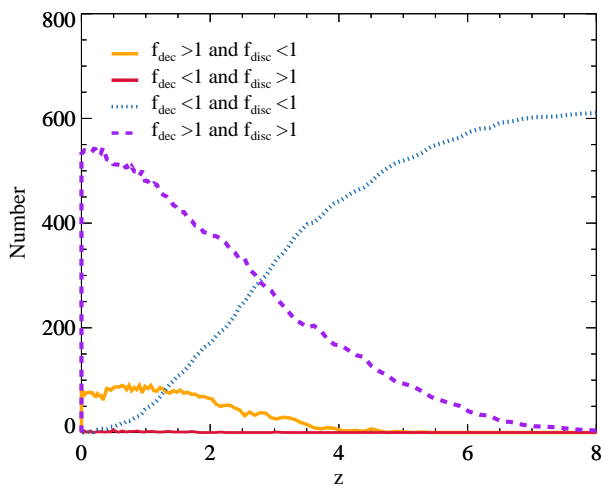}
\caption{Evolution of the number of MW-like galaxies with different combinations
of disc instability criteria.
The orange line indicates the number of galaxies that fulfill
both criteria to allow bar instability in the disc. The blue dotted line shows 
the number
of galaxies that satisfy the ELN criterion, but not the 
\citet{Algorry2017} criterion. The purple 
dashed line shows the number of galaxies in the opposite situation, where the 
\citet{Algorry2017} criterion is satisfied, but the ELN is not. }
\label{fig:InsMW}
\end{figure}

The impact of the criteria used to determine DI in MW-like galaxies
is presented in Fig.~\ref{fig:InsMW}.
The orange line shows
the number of MW-like 
galaxies that suffer disc instabilities as a function of redshift. The 
other lines in the plot show the number of galaxies that fulfill only one of the 
criteria to allow instability of the disc, but not the other.
The blue dotted line indicates the number of galaxies with $f_{\rm dec} < 1$ and 
$\epsilon_{\rm thresh} < 1$ and the purple dashed line shows the number of galaxies 
with $f_{\rm dec} > 1$, but $\epsilon_{\rm thresh} > 1$. We can see that instability events
in MW-like galaxies in our simulation start roughly at $z\sim4$. The number of 
galaxies suffering DI increases with time, reaching a maximum at approximately $z\sim0.5$
and declines slightly towards $z=0$. This is in line with the current paradigm of 
galactic evolution where, at early stages, galaxies form rapidly through 
hierarchical clustering and mergers, while at later times, secular transformations drive
the evolution of galaxies, frequently drived by bar instabilities \citep{Kormendy2013}.

\section{Results}
\label{sec:results}

In this section, we analyze the properties of the SPs of MW-like galaxies 
selected combining the two criteria presented in Section~\ref{sec:MW}. 
We concentrate in the IDFs of the SPs and their abundances of alpha-elements.  
  
\subsection{Metallicity distribution in simulated bulges}
\label{sec:IDF}

\citet[][ Z08 hereafter]{Zoccali2008} analyzed the data of 800 bulge field 
K giants stars observed in the VVV survey in three different 
latitudes ($b=4\degree$ , $b=6\degree$ and $b=12\degree$) and constructed 
the iron distribution function (IDF) based on    
medium ({\rm FLAMES-GIRAFFE}) and high resolution spectra ({\rm UVES}). 
The Vista Variables in the V\'ia L\'actea 
survey (VVV) provides state of the art photometric 
observations of more than $10^6$ stars in the MW bulge. 
We construct an IDF making use of 
the metallicity of $1297758$ stars, computed by means of the method described in 
\citet[][ section 4]{Gonzalez2013}. Metallicities are derived by interpolating the 
colors of individual red giant branch stars between globular cluster ridge lines 
with well known metallicities. We compare the results obtained from {\rm SAG} with 
both observational IDFs. 
Each one has advantages and drawbacks when comparing with the model IDF. 
On one hand, the metallicities obtained from the spectroscopic survey are
more fairly comparable with metallicities given by the model. Metallicities 
provided by the photometry are a more indirect method. On the other side, the 
spectroscopic IDF does not reflect the number density of stars in each
latitude of the bulge in which the observations were made.  Stars of the spectroscopic survey
were observed in three fields along the Galactic minor axis with a similar number of 
stars ($204$ at Baade's window, $213$ at $b=-6\degree$ and $104$ at $b=-12\degree$).
This means that stars with high metallicity in the spectroscopic sample are underrepresented, 
considering that the density of stars near the galactic plane is larger than at high latitudes. 
For this reason, we compare the model IDF with both photometric and spectroscopic IDF.

\begin{figure}
   \includegraphics[scale=0.37]{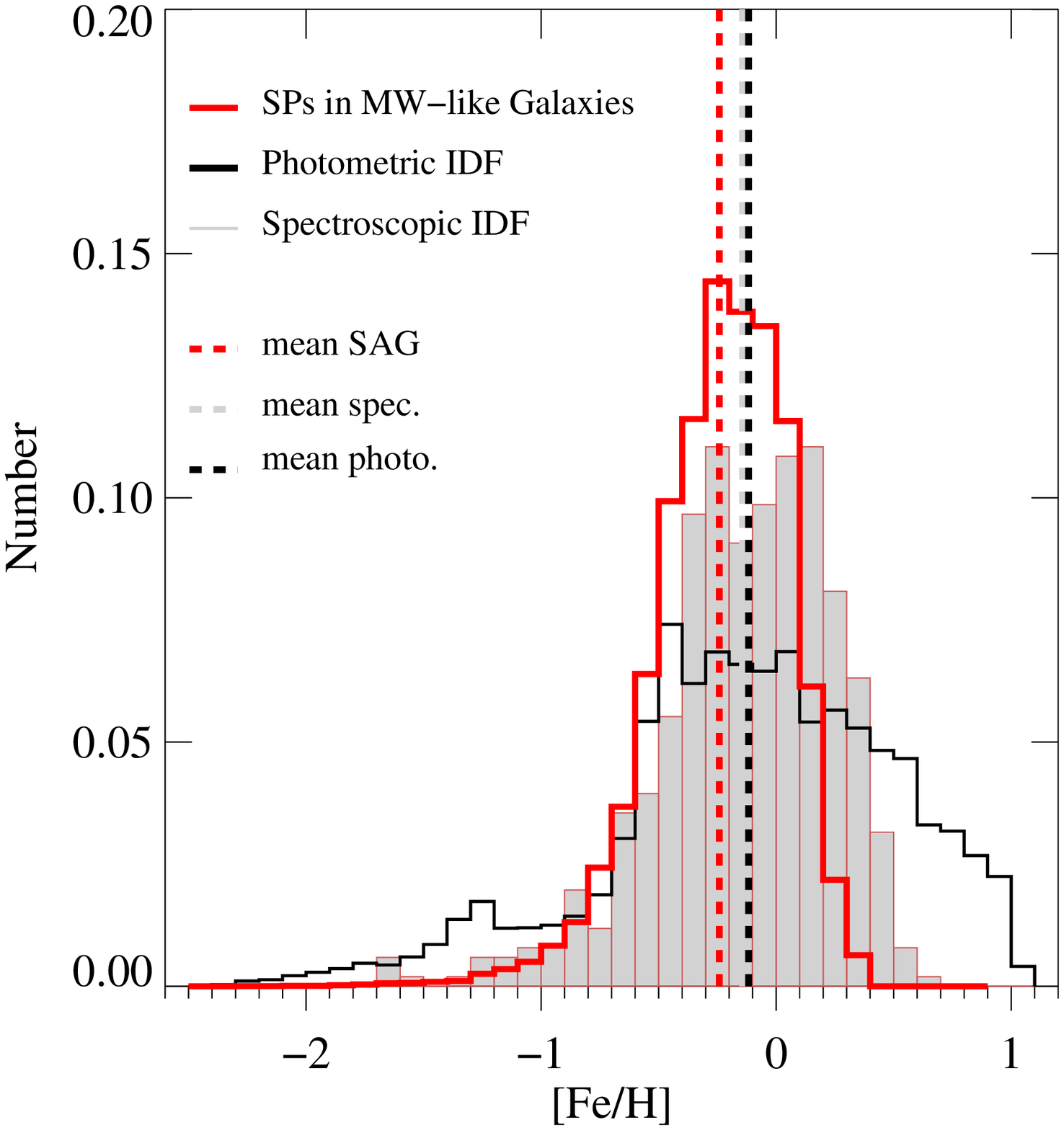}
   \includegraphics[scale=0.37]{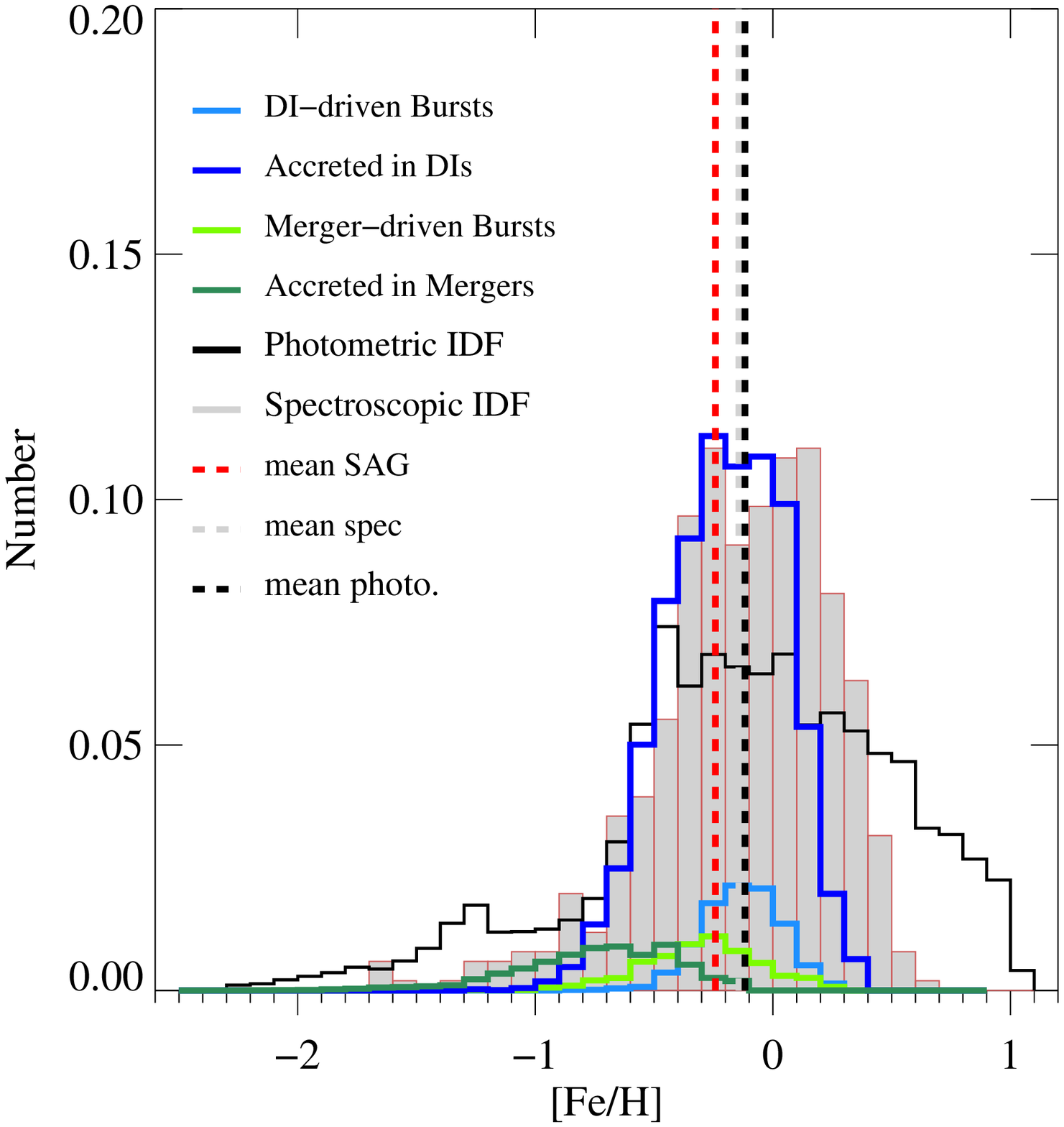}
\caption{{\em Upper panel:} Metallicity distribution of all the 
SPs in bulges of MW-type galaxies in the model (red histogram). For
comparison, the spectroscopic IDF of \citet{Zoccali2008} (grey 
histogram) and the photometric IDF constructed for this paper are shown. Dashed and 
dashed-dotted lines represent the mean values of the 
model and observed distributions, respectively.
{\em Lower panel: }
Same distributions as in the upper panel, but discriminating the
SPs according to their origin, that is, to the process involved in bulge formation.
The blue line represents 
the SPs accreted in disc-instabilities (DIs), while the light-blue line corresponds the SPs 
formed in burst during DIs. Light-green line depicts the SPs formed in bursts
during mergers and the green line shows the SPs accreted in mergers. 
}
\label{fig:SP-IDF}
\end{figure}

Fig.~\ref{fig:SP-IDF} shows the IDF of SPs of the bulge
of MW-like galaxies selected in our simulation. 
The upper panel shows the normalized total iron distribution from the 
model (red histogram outline), compared to the spectroscopic IDF from 
Z08 (grey histogram) and the photometric sample (black histogram outline). The mean 
metallicity obtained for the simulated bulge
SPs is $[\text{Fe/H}] = -0.24$, while the mean 
metallicity of the spectroscopic IDF is $[\text{Fe/H}] = -0.14 $. The photometric IDF has the same 
mean metallicity than the spectroscopic one, rounded to the second decimal place. These
values are highlighted with vertical dotted lines in the plot. 

The general shape of the model IDF resembles the spectroscopic one. 
The model IDF lack SPs with metallicities 
higher than $0.5$ dex above the average and show a sharp cutoff, while the spectroscopic IDF 
shows stars with 
$[\text{Fe/H}]$ higher than $0.7$ dex from the average and also a relatively sharp cutoff. On the other 
hand, the photometric IDF, with its natural correct weight given to the stars with higher 
metallicities, extends to $1.2$ dex higher than the average. 
The low metallicity tail shown by both observed metallicity distributions
is recovered by our model. 

The lower panel of Fig.~\ref{fig:SP-IDF} shows 
the IDF in bulges of MW-type galaxies 
compared to the observed IDFs but
 SPs originated 
in the different processes involved in bulge formation are distinguished. 
IDF of SPs formed in bursts during DIs and accreted in DIs are represented by 
light-blue and blue histograms, respectively, while the light-green histogram
depicts the SPs formed in bursts
during mergers and the green histogram shows the SPs accreted in mergers.
We can see that DI is the main physical process contributing with SPs that
form the bulge of MW-type galaxies. 
They represent, on average, $87\%$ percent of the total mass of the stacked bulge
of MW-type galaxies. More precisely, the stars driven to the bulge due to disc
instabilities contribute with $78\%$ percent of the bulge mass, and the  
starbursts during disc instabilities are responsible for $9\%$ percent of 
the final mass. The contribution of mergers 
is of roughly $13\%$ percent of the final
bulge mass, with a share of $7\%$ percent of stars accreted in mergers and 
$6\%$ percent of stars formed during starbursts in mergers. Although mergers 
are responsible for a minor fraction of the mass of bulges in model galaxies, 
stars accreted in mergers allow to produce the low metallicity tail of the IDF,
recovering the observed trend.  

The stars formed in bursts, either triggered by mergers or by disc 
instabilities, present higher metallicities on average than the stars accreted 
into the bulge. This is due to the extended bursts that are taken into account 
in \sag; the new generations of stars are formed from 
cold gas available for bulge formation that is consumed gradually
while it is contaminated by successive star formation events.

\subsection{Alpha-element abundances of SPs}

The $\alpha$-element abundances of stars in the bulge of the MW 
are source of valuable information when we are studying the 
formation history of this component.

The abundance ratios
of these elements with respect to iron are well known as indicators
of formation time scales of a SP \citep{Tinsley1979}. 
SNe CC
polute the ISM with elements 
like  Mg, O, S, Si and Ca 
($\alpha$-elements)
, while SNe Ia provide
iron. 
Since SNe CC have much shorter lives 
than SNe Ia ($\approx 0.1 - 3 \,{\rm Gyr}$)
a SP with high abundance of Mg, O, S, Si, or Ca with respect to Fe is thought
to be formed rapidly, from gas 
still not contaminated with the 
yields of SNe Ia explosions. 
Different works have addressed the abundance of  
$\alpha$-elements in stars in the bulge of the MW. 
Analyzing a K giant sample in the bulge, \citet{McWilliamRich1994} found
overabundance of $\alpha$-elements with respect to solar abundances
for Mg and Ti for the whole measured interval of $[\text{Fe/H}]$, while 
Ca and Si abundances showed solar abundances for $[\text{Fe/H}] > -0.2$, 
resembling disc stars. They discuss that the excess of 
$\alpha$-elements can reflect an enrichment process in common 
with elliptical galaxies. 
\citet{Zoccali2006} measured the oxygen
abundances of 50 K giant 
stars in the MW bulge and found that they have $[\text{O/Fe}]$ ratios  
higher than solar and higher than the stars in the thick disc, which, 
in turn, show higher $\text{[O/Fe]}$ than the stars in the thin 
disc \citep{Bensby2004}. They also consider 
that the 
evidence found gives support to the vision of the bulge as an old 
spheroid, with short formation time scales, as a 
classical bulge. 

\citet[][G11 hereafter]{Gonzalez2011} study the abundance ratios   
$[\alpha/\text{Fe}]$
of 474 stars in the giant branch of the MW bulge and 176 more 
belonging to NGC 6553 using the FLAMES-GIRAFFE instrument in VLT;
the sample is the same used in Z08. At the same time, they 
reanalyzed the sample of stars of the thick and thin disc from 
\citet{AlvesBrito2010} and obtained abundances from these components
in an homogeneous way
to study the origin of the different components of the galaxy. 
They found that the stars with high metallicities have low 
$[\alpha/\text{Fe}]$ 
ratios and are found exclusively at lower latitudes ($b=-4\degree$),
while stars of low metallicity are 
evenly spaced in the
bulge and have identical properties with the stars of the thick disc.
They also found that the population of stars of low metallicity 
have 
oversolar 
$[\alpha/\text{Fe}]$
abundance ratios, higher than the stars
in the thin disc, and conclude that the origin of this population
is consistent with a clasical bulge, with short formation timescales.
G11 also found a 
metallicity gradient, with higher metallicities
near the galactic plane.
However, they do not detect a gradient in $[\alpha/\text{Fe}]$, 
as would be expected if two fully distinct 
populations exist characterized by different formation histories and timescales.

\begin{figure*}
   \includegraphics[scale=0.35]{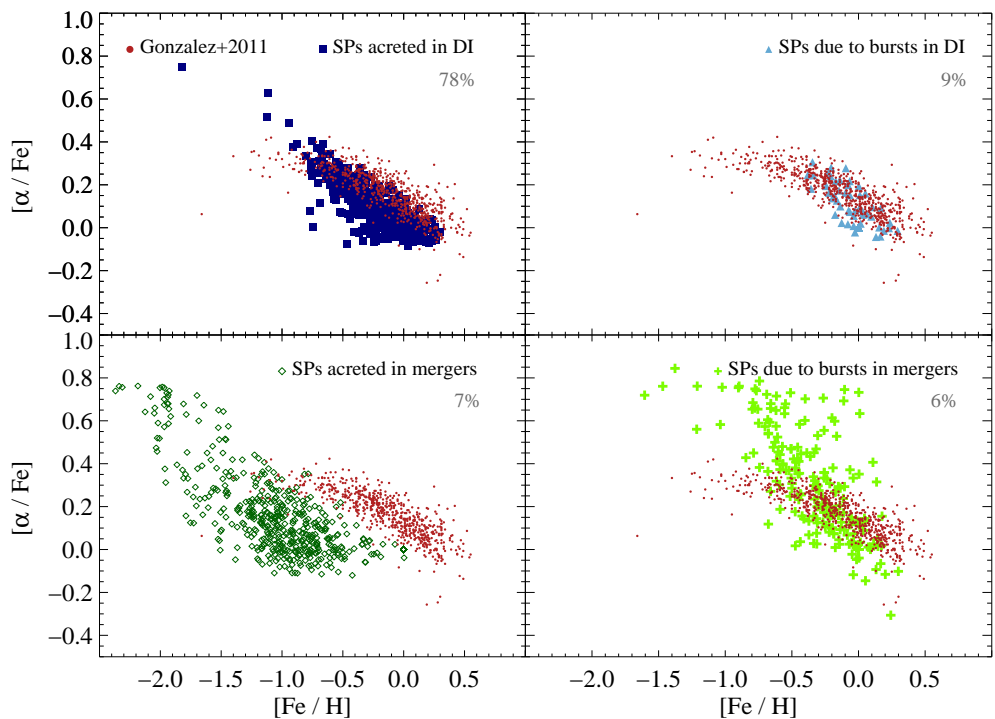}
\caption{Alpha element abundances of bulge SPs 
of MW-like galaxies in $\sag$. Observations of stars in the bulge of the
MW from \citep{Gonzalez2011} are indicated with red dots. In each panel,
different formation mechanisms of SPs are shown. In grey, we note the 
contribution
to the total mass of the bulge by each mechanism. {\it Upper left}: SPs
accreted in DIs. {\it Upper Right: }  SPs originated in bursts due to DIs.
{\it Lower left:} SPs accreted in mergers {\it Lower Right: } SPs originated
in bursts during mergers. }
\label{fig:figAFe}
\end{figure*}

We construct the $\alpha$-element abundances of SPs 
in our model by suming the abundance ratios [\text{Mg/Fe}], [\text{Si/Fe}] 
and [\text{Ca/Fe}] as presented in G11.
Fig~\ref{fig:figAFe} 
shows the [$\alpha$/\text{Fe}] ratios, 
vs $[\text{Fe/H}]$, which are predictions of our model. In each panel,
SPs with different origins are shown.The lower average iron abundances in 
SPs of MW-like galaxies seen in Fig.~\ref{fig:SP-IDF} is naturally present 
here.
We can see that the [$\alpha$/\text{Fe}] ratios
of SPs are, in general, in good agreement with observations, presented in
red dots in all the panels, although the SPs originated in mergers reach values
of [$\alpha$/\text{Fe}] ratios above the plateau  
developed by the observations. Moreover, SPs accreted in mergers  
of satellite galaxies, are offset low by more than $0.5$ dex in $[\text{Fe/H}]$,
while the IDF is offset only by $0.05$ dex with respect to observations. The 
behavior of these SPs is clearly different than SPs originated in-{\it situ}. 
However, although we find a large number of SPs accreted in mergers, they   
contribute with only $7\%$ of the mass to the bulges of the MW-like galaxies. 
The likelihood to observe stars with properties similar to those of accreted
SPs in the model is low and hence they can be less conspicuous in the observed
data. Besides, these SPs come from stellar systems with different properties 
between them and the analysis of their chemical enrichment is not 
straighforward.   
The slopes of
the trends of [$\alpha$/\text{Fe}] with $[\text{Fe/H}]$  are in good agreement
with the data from G11. However, the model abundances of SPs do not show 
a knee at low 
metallicities, irrespective of their origin. This could indicate that SN Ia are
contributing with iron too early in our simulation. \citet{Gargiulo2015} 
showed in their Fig. $8$ the delay times of explotion of SN CC and 
SN Ia used in this article. There is a peak of SN Ia near 0.1 Gyrs, a slow
decline up to 2 Gyrs and then, the distribution becomes steep up to 10 Gyrs. 
 The SN Ia delay time 
distribution (DTD) of SNIa is a matter of debate. \citet[][ GR08 afterwards]{Greggio2008} compare DTDs modeled with different assumptions regarding the degree of orbital shrinkage during the first common envelope phase \citep{Greggio2005}. A SN Ia delay time 
distribution more similar to the DD WIDE case shown in Fig. 1 of GR08 
could contribute to the development of a knee in 
the [$\alpha$/\text{Fe}]-$[\text{Fe/H}]$ space, since a larger number of SNIa
would contribute to the enrichment in later phases of the chemical evolution of
the SPs that turn out to be in the bulge.

 SPs that form the bulge through DI events
seem to contribute with stars in the whole range
of observed [\text{Fe/H}] and [$\alpha$/\text{Fe}]. 
This fact supports    
the lack of gradient of [$\alpha$/\text{Fe}] abundances with latitude
in stars along the minor axis as found by G11. Moreover, SPs in our model
 generated in DIs reach lower values of [$\alpha$/\text{Fe}] abundances than 
SPs formed and accreted in mergers. This is also in agreement with the results
 of G11, in the sense that low metallicity SPs, like SPs originated in mergers 
in our model, have higher [$\alpha$/\text{Fe}] abundances and are 
thought to be part of a classical bulge component. 

\section{Discussion and Conclusions}
\label{sec:discandconc}

We use the semi-analytic model of galaxy formation \sag, run on top the 
cosmological {\em N}-body simulation MDPL2,
part of the MULTIDARK project 
(See Section~\ref{sec:model}), to study the formation and chemical 
enrichment of stars in bulges of MW-like galaxies,
with the aim of contributing to our understanding of the formation
scenario of MW-like galaxies.
A novel approach to follow up SPs in the semi-analytic model is introduced. 
We look for galaxies in the model that resemble the MW. 
For this purpose, we apply two different criteria 
simultaneously, one of them
accounting for the cosmological context of the local group
in which a MW-like galaxy might reside,
and the other for the morphological classification
of MW-like galaxies. Fig.~\ref{fig:SPLGM11} 
shows that in our model a galaxy with a bulge
resembling that of the MW
is a particular case among galaxies that reside in 
DM halos that share the same cosmological context 
as the MW halo. Galaxies
in LG analogs tend to have more massive bulges.
On the other hand the SFR histories shown in Fig.~\ref{fig:SfrMW}
indicate that galaxies with the same cosmological context
as LG galaxies and with the MW morphology form stars less
efficiently than galaxies selected by any of the criteria
separately.
Thus, as inferred from our model,
galaxies residing in LG analogs
become similar to the MW 
if they are inefficient forming 
stars and building up a massive bulge, or completely relaxing 
into an elliptical galaxy. This also indicates that considering 
a single criterion to select MW-like galaxies in simulations
could result misleading.
\\
We introduce a novel modelization of disc instability events in \sag.
The criticized ELN criterion is combined with an extra constraint that
takes into account the gravitational importance of galactic discs
in relation with their host DM halos, following the results obtained by
\citet{Algorry2017} (see Section~\ref{sec:DI}). When the disc becomes gravitationally
dominant of the whole system, it becomes suitable for an instability event. 
If in addition to this
the galaxy is perturbed by a neighboring galaxy, then it becomes
unstable. We also include a gradual transfer of mass to the bulge due to 
disc instabilities. 
The combined effect of these implementation allows to
obtain a population of MW-like galaxies 
with bulge properties which are in good agreement with observations,
overcoming one of the major drawbacks in the modeling of morphological evolution 
of galaxies in semi-analytic models.
\\
A calibrated version of the model using general 
constraints of observed galaxies 
predicts an iron metallicity distribution of
SPs with a shape that is in good agreement with spectroscopic 
observations, although the 
average metallicity of the modeled SPs is offset slightly low 
by $\approx 0.05$ dex with respect to the observed distribution 
presented by Z08 and the photometric IDF constructed for this article.  
In addition, if we consider the comparison of our model IDF with the
photometric IDF, the lack of SPs with high metallicity in the model is more 
evident. 
This underestimation can be interpreted considering
the way in which the different baryonic components are polluted in the model.
Semi-analytic models are conceived to reproduce global 
properties of galaxies and their components. Stars contaminate the gas phase
by the stellar yields ejected at the end of their lives.
A large amount of the recycled material ends in the hot 
gas of the galaxy host halo, being strongly diluted. Part
of this hot gas can be transferred to an ejected reservoir
and later reincorporated to the hot phase keeping the same
 low mean metallicity. Therefore the metallicity of the cold
gas disc is lower than expected, because it is poluted
 by gas cooling of the hot phase. This issue could be solved 
if our semi-analytic model considered density profiles for the 
different baryonic components (gas and stars), allowing the development 
of metallicity gradients. In the case of the hot gas, the cooling flows 
generated in the inner
regions could contribute to the cold gas with higher levels of metallicity, 
increasing the mean
metallicity of the whole galaxy. Besides, SPs formed at lower  galactocentric
distances would have higher levels of metallicities, helping to 
avoid the sharp cutoff in the model IDF.\\

When we discriminate the origin of the different SPs in our models, we found 
that the main driver of bulge formation in MW-like galaxies are 
disc instabilities which give place to 
transfer of mass from the disc and the formation of new stars.  
These stars contribute to
the range of intermediate metallicities and populate the more 
metallic 
region of the IDF. This seems to agree with observations, since \citet{Ness2013}
conclude that the more  
metallic stars in the MDF observed with ARGOS 
spectroscopic survey are transferred to the bulge by disc instability events.
We also show that there 
is a non-negligible fraction of stars, about $13\%$ of the total SPs of the 
stacked bulge in our simulation, 
that, on one hand, are accreted in mergers of 
low mass galaxies, and on the other, are formed in bursts due to the violent
relaxation generated by mergers.
Those subpopulations, although less abundant, 
might explain the low metallicity tail of the observed
IDF. This is in 
qualitative agreement with the characterization of the metal poor component
of the MDF of stars in the bulge
done by \citet{Zoccali2016}, who point out that this component, although 
present at high latitudes, might not be very conspicuous.
These arguments are reinforced when we study the [\text{$\alpha/\text{Fe}$}] 
abundance ratios of SPs in bulges of MW-type galaxies in our model. The possibility 
of the existence of two different groups of SPs in the observed bulge is again 
revealed. 
SPs originated in mergers in our model, 
with lower metallicities reach higher [\text{$\alpha/\text{Fe}$}] values than SPs
originated in DIs. Besides, SPs originated in DIs populate the whole range of 
[\text{$\alpha/\text{Fe}$}] that span the observations, explaining the observed lack 
of a gradient of [\text{$\alpha/\text{Fe}$}] abundance ratios along the minor axis 
in the bulge.

\section*{Acknowledgments}

This work was partially supported by the Consejo Nacional de Investigaciones Cient\'{\i}ficas y
T\'ecnicas (CONICET, Argentina), 
Universidad Nacional de La Plata (UNLP, Argentina), 
Instituto de Astrof\'{\i}sica de
La Plata (IALP, Argentina) and 
Secretar\'{\i}a de Ciencia y Tecnolog\'{\i}a de la Universidad
Nacional de C\'ordoba (SeCyT-UNC, Argentina).
IDG acknowledges support from  Comisi\'on Nacional de Investigaci\'on Cient\'{\i}fica y Tecnol\'ogica (CONICYT, Chile) and Pontificia Universidad Cat\'olica de Chile (PUC) projects PFB-06 and ACT-86 for an academic stay at PUC.
SAC acknowledges grants from CONICET (PIP-0387), Argentina,
Agencia Nacional
de Promoci\'on Cient\'{\i}fica y Tecnol\'ogica (PICT-2008-0627), Argentina,
and Fondecyt, Chile.

The authors gratefully acknowledge the Gauss Centre for Supercomputing e.V. (www.gauss-centre.eu) and the Partnership for Advanced Supercomputing in Europe (PRACE, www.prace-ri.eu) for funding the MultiDark simulation project by providing computing time on the GCS Supercomputer SuperMUC at Leibniz Supercomputing Centre (LRZ, www.lrz.de).
The CosmoSim database used in this paper is a service by the 
Leibniz-Institute for Astrophysics Potsdam (AIP).
The MultiDark database was developed in cooperation with the 
Spanish MultiDark Consolider Project CSD2009-00064.
Part of the analysis was done on the Geryon cluster at the Centre 
for Astro- Engineering UC. The Anillo ACT-86, FONDEQUIP AIC-57,  
Newton-CONICYT fund project DPI20140114 and QUIMAL 130008 provided 
funding for several improvements to the Geryon cluster. 

\newpage
\bibliographystyle{mn2e}
\bibliography{gargiulo-SPs}

\bsp

\label{lastpage}

\end{document}